December 2023

# Personalized Content Moderation and Emergent Outcomes

Necdet Gürkan
*Stevens Institute of Technology*, ngurkan@stevens.edu

Mohammed Almarzouq
*Kuwait University*, mo.almarzouq@ku.edu.kw

Pon Rahul Murugaraj
*Indian Institute of Management Bangalore*, ponrahul.m21@iimb.ac.in

# Personalized Content Moderation and Emergent Outcomes

*Short Paper*

**Necdet Gürkan**
School of Business
Stevens Institute of Technology
ngurkan@stevens.edu

**Mohammed Almarzouq**
Department of Information Systems and Operations Management, Kuwait University
mo.almarzouq@ku.edu.kw

**Pon Rahul Murugaraj**
Indian Institute of Management Bangalore
ponrahul.m21@iimb.ac.in

## Abstract

*Social media platforms have implemented automated content moderation tools to preserve community norms and mitigate online hate and harassment. Recently, these platforms have started to offer Personalized Content Moderation (PCM), granting users control over moderation settings or aligning algorithms with individual user preferences. While PCM addresses the limitations of the one-size-fits-all approach and enhances user experiences, it may also impact emergent outcomes on social media platforms. Our study reveals that PCM leads to asymmetric information loss (AIL), potentially impeding the development of a shared understanding among users, crucial for healthy community dynamics. We further demonstrate that PCM tools could foster the creation of echo chambers and filter bubbles, resulting in increased community polarization. Our research is the first to identify AIL as a consequence of PCM and to highlight its potential negative impacts on online communities.*

**Keywords**: personalized content moderation, social media platforms, asymmetric information loss, echo chambers

## Introduction

Social media platforms, including Facebook, Instagram, and Reddit, play an increasingly significant role in civic engagement, serving as essential forums for discourse. These platforms collectively provide a space for the public to gather, discuss, debate, and share information, a role actively promoted through their messaging. However, as these virtual spaces facilitate the exchange of information, online harms such as hate speech and harassment have become major threats to internet users (Thomas et al., 2021). In response, online platforms have developed automated tools to combat toxic content, such as Instagram's harassment comment detector (Instagram, 2019), Google's Jigsaw API for detecting toxic comments (Jigsaw, 2017; Wulczyn et al., 2017), and Yahoo's abusive language classifier (Nobata et al., 2016).

Although social media platforms have employed various content moderation tools to enforce their usage policies, there exists a nuanced spectrum of content that, though not sufficiently toxic to explicitly violate these policies, may still wield a detrimental impact on users (International, 2020). Additionally, content deemed toxic may sometimes restrict users' freedom of speech, presenting a complex challenge in balancing safety and expression. The "gray areas" in determining what constitutes toxic content online arise from divergent user opinions. These differences are often rooted in individual lived experiences, cultural perspectives, political views towards free speech, or varying access to appropriate context (Gualdo et al., 2015; Sambasivan et al., 2019). Given individual differences, a one-size-fits-all solution moderation mechanism is not suitable for serving the disparate needs of millions of end users (Jiang et al., 2023).

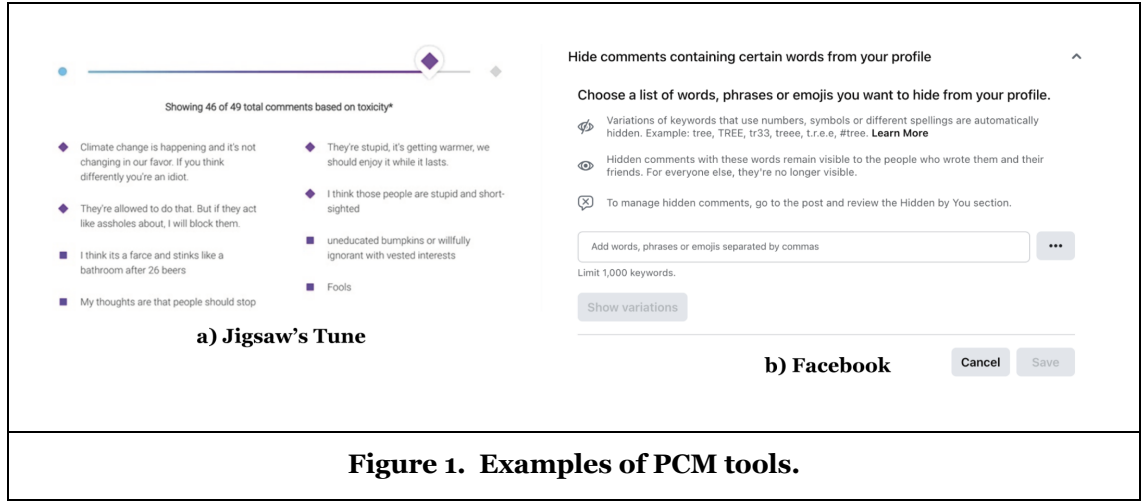

**Figure 1. Examples of PCM tools.**

Recognizing the complexities in centralized content moderation, some industry leaders, scholars, and activists advocate for Personal Content Moderation (PCM), which places moderation controls and preferences in the hands of users or aligns its algorithms to each user's preferences (Jhaver et al., 2023). Recently, many platforms have begun experimenting with such tools, providing users with greater control over their experiences. These tools are 'personal' because each user can configure them differently, and a user's configuration only applies to their own account. Examples of PCM tools include toggles, sliders, or scales for attributes like 'toxicity' or 'sensitivity', as well as word filter tools that block user-specified phrases, as shown in Figure 1. Researchers have also begun to develop adaptive PCM algorithms, which automate this process by learning from users' previous behaviors and preferences (Jhaver et al., 2022; Kumar et al., 2021). While traditional account-based personal moderation tools allow users to block or mute content from specific accounts, PCM tools perform moderation based on the content of the messages themselves (Geiger, 2016).

| **High-sensitivity view** | **Low-sensitivity view** |
|---|---|
| "*I think people who received COVID-19 vaccination are beefheaded... Clearly, you did not see the recent studies! [LINK]*" | [deleted] |
| "*It makes sense. I don't know why nobody is talking about it.*" | "*It makes sense. I don't know why nobody is talking about it.*" |
| "*Stop sharing low-quality studies*" | "*Stop sharing low-quality studies*" |
| **Table 1. Example discussion thread on COVID-19 vaccination.** ||

A concern that might be easily overlooked with the increasing adoption of PCM is the potential for personalized moderation to create divergent views of the same discussion post. In certain instances, this could lead to a loss of meaning for some users. In more severe cases, it could even become a source of misunderstanding and conflict. We describe this phenomenon of variable information loss as 'asymmetric information loss' (AIL), as illustrated in Table 1.

In conventional one-size-fits-all moderation schemes, asymmetric information loss (AIL) was less apparent, as moderated content remained consistent for all users, thus preventing such disparities. However, PCM, while potentially enhancing individual experiences, could have detrimental long-term effects on community dynamics due to AIL. AIL occurs when there is variability in information loss experienced by users, shaped by their personal preferences in PCM settings.

Moreover, reliance on PCM tools may promote the formation of echo chambers (Sunstein, 2001) and filter bubbles (Pariser, 2011), leading to increased community polarization. Echo chambers occur when individuals are exposed primarily to opinions and information that align with their own, reinforcing pre-existing views. Filter bubbles, on the other hand, are algorithmically curated content environments that isolate users from a broader range of perspectives. Both phenomena limit exposure to diverse viewpoints, potentially intensifying polarization. These potential outcomes highlight the urgency of investigating whether PCM can lead to negative emergent effects in online communities. Therefore, we propose our research question as follows:

**RQ:** How does PCM affect emergent outcomes in social media discussions?

## Methods

Reddit is a social media platform where users share posts and comment on them. It is organized into numerous subgroups — known as *subreddits*—each covering a wide range of topics, such as *r/television*, *r/askscience*, and *r/cryptocurrency*. This study focuses on two specific communities: *r/mentalhealth* and *r/politicaldiscussion*. The *r/mentalhealth* subreddit is a space for discussing, venting, supporting, and sharing information about mental health, illness, and wellness. The *r/politicaldiscussion* encourages users to have substantial and civil political discussions. The selection of these two communities was guided by the hypothesis that they might exhibit varying levels of tolerance for toxicity among their users, thereby allowing us to observe the impact of PCM on individuals with different degrees of sensitivity.

We compiled a dataset comprising the 50 most discussed posts from each of the *r/mentalhealth* and *r/politicaldiscussion* subreddits, resulting in a total of 100 posts. These posts were specifically selected based on their high comment counts, thereby capturing the most actively engaged topics within each community. This data was obtained from the Pushshift Reddit dataset (Baumgartner et al., 2020). For each of these subreddits, we included posts made between June 1, 2022, and January 1, 2023. Within the *r/mentalhealth* subreddit, the most discussed 50 posts had a total of 12,737 comments, with the comment count per post ranging from 199 to 520. Within the *r/politicaldiscussion* subreddit, the most discussed 50 posts had a total of 13,658 comments, with the comment count per post ranging from 75 to 1374.

### *Measures*

**Perspective API**

For the measurement of content toxicity in online discussions, our study employed the Perspective API, a machine learning-based tool developed by Jigsaw, a subsidiary of Google (Jigsaw, 2017). This advanced tool analyzes text inputs, such as social media comments or discussion forum posts, and assigns a toxicity score ranging from 0 to 1. This score reflects the likelihood that an average human reader would perceive the content as toxic, with 0 indicating no perceived toxicity and 1 indicating a high likelihood of being perceived as toxic. The API's algorithm evaluates various factors, including the usage of harmful language and the context of words or phrases, to generate this score. We applied the Perspective API to calculate the toxicity score of all comments in our sample. Subsequently, we used this standardized score to simulate different levels of user sensitivities by excluding comments that met various toxicity thresholds, thus representing the assumed sensitivities of our users.

**Asymmetric Information Loss (AIL)**

*Asymmetric information loss* (AIL) refers to the variability in information loss experienced by users due to personalized moderation activities (Table 1). To operationalize this concept, we quantified the loss of information for an individual by measuring the degree of similarity—or lack thereof—between the moderated discussion post observed by the user and the corresponding non-moderated discussion post. We utilized cosine similarity from RoBERTa text embeddings (Liu et al., 2019) to quantify the difference between unmoderated discussion posts and different moderated versions of it in which comments were removed based on varying degrees of toxicity thresholds ranging from 0.01 to 0.99. Utilizing embeddings to extract semantic and social knowledge for understanding various phenomena has been applied in a wide range of studies (Chang et al., 2023; Gurkan & Yan, 2023). This method was applied to 50 different discussion posts, enabling us to determine the information loss at each toxicity tolerance threshold.

**Polarization**

Echo chambers are a key indicator of polarization (Sunstein, 2001). Echo chambers, in principle, suggest that a person is not being exposed to views that differ from their own (Garimella et al., 2018). This effect is amplified in online communities, where algorithms, user preferences, and moderation often create a self-reinforcing cycle of shared beliefs and opinions. Within these echo chambers, the shared reality becomes homogenized, as members are predominantly exposed to information that aligns with their existing viewpoints. Our semantic analysis aims to estimate this shared reality by examining how PCM influences language usage.

As an initial step, we employed various NLP techniques using Python's Natural Language Toolkit (NLTK) (Bird et al., 2009). These included removing common words (stop words), converting words to their base form (lemmatization), transforming text to lowercase, and eliminating punctuation. Subsequently, we identified the top 100 words in each dataset based on their frequency, that is, the number of times each word appears. This limitation to 100 words was aimed at focusing on the most prevalent words while excluding those less commonly used. This approach of concentrating on the top 100 most frequently used words aligns with methodologies employed in other language usage research studies (Ibrahim et al., 2017; Lyddy et al., 2014).

We then used the Jaccard index to quantify the degree of overlap in information dissemination to analyze the shared reality within each community (Erickson et al., 2023). The Jaccard index is a commonly used similarity metric in information retrieval for computing the similarity between two sets (Niwattanakul et al., 2013). In this study, the Jaccard index measures the similarity between the most frequently used words in two scenarios: comments that remain after PCM and comments as they are without any PCM. We methodically adjusted the Perspective API's toxicity score threshold, ranging from 0.01 to 0.99, for each post comment. At each of these threshold levels, we calculated the Jaccard index. This helped us quantify the similarity between the sets of original comments and those that would be removed if they exceeded the specified toxicity threshold. We used this measure as a proxy for polarization, based on the rationale that a high similarity indicates PCM has minimal impact on the words used in discussion posts. Conversely, a significant difference suggests that PCM is likely restricting the variety of ideas shared within the community, effectively turning it into an echo chamber and, as a result, leading to polarization.

## Results

The results from the Perspective API revealed similar levels of toxic language in the *r/mentalhealth* ($M$ = 0.22, $SD$ = 0.20) and *r/politicaldiscussion* ($M$ = 0.24, $SD$ = 0.16) subreddits. This finding was unexpected, particularly because we had initially hypothesized that *r/politicaldiscussion* would have higher toxicity levels, given the often-heated nature of political discourse, as opposed to the *r/mentalhealth* community. However, the relatively low toxicity score in *r/politicaldiscussion* might indicate rigorous moderation to control toxic language. This suggests the presence of strong community norms influencing language use as consistent with previous findings (Erickson et al., 2023; Gurkan & Suchow, 2022). Additionally, Figure 2a highlights two key aspects: it illustrates the correlation between content removal frequency and the varying PCM thresholds as set by the Perspective API, and it also shows that the *r/mentalhealth* and r/*politicaldiscussion* communities have similar rates of content removal at these thresholds.

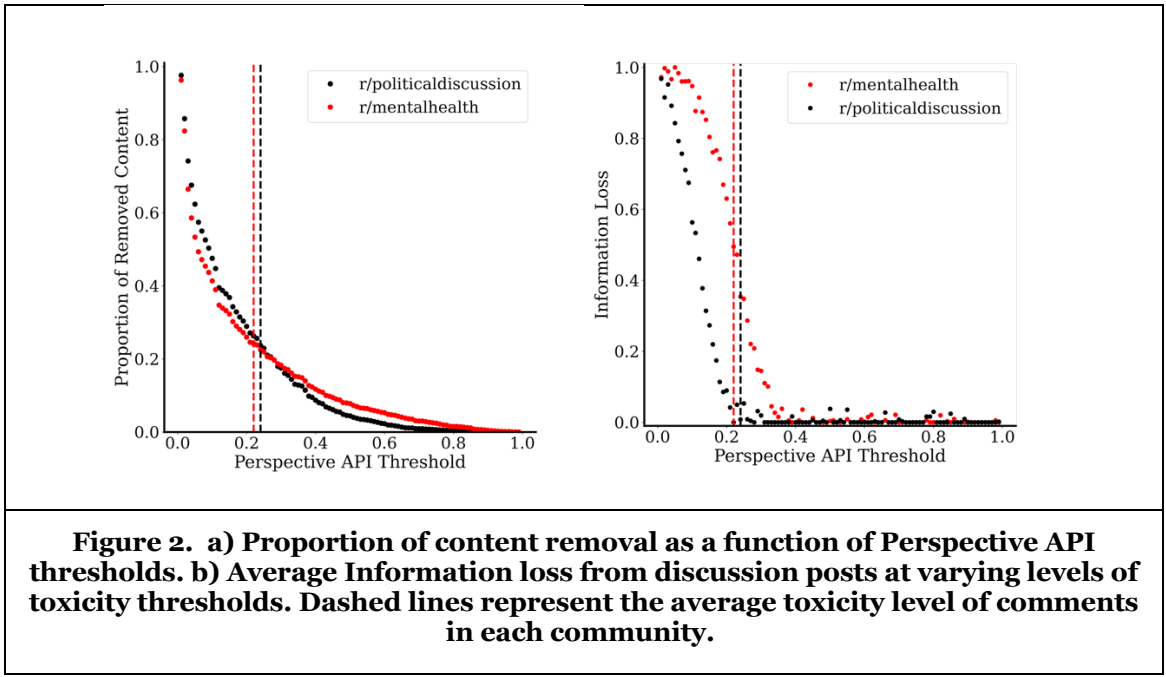

**Figure 2. a) Proportion of content removal as a function of Perspective API thresholds. b) Average Information loss from discussion posts at varying levels of toxicity thresholds. Dashed lines represent the average toxicity level of comments in each community.**

To assess whether the PCM results in *AIL* within the community, we examined the cosine similarity of embeddings between moderated and unmoderated discussions at varying levels of toxicity thresholds. Figure 2b demonstrates that PCM implementation leads to varying degrees of information loss for users, contingent upon their individual toxicity threshold settings. Unlike a one-size-fits-all approach, which would cause a consistent level of information loss across all users, PCM's personalized approach results in differing impacts. Notably, users with toxicity thresholds below the community average in the *r/mentalhealth* subreddit are likely to experience more substantial information loss compared to those with higher thresholds. This effect is especially significant in the *r/mentalhealth* community, likely due to its emphasis on information exchange, potentially causing greater AIL with PCM.

We compared the most frequently used words in comments before and after PCM implementation to evaluate how PCM alters community discourse and creates echo chambers. By analyzing these differences, reflected in the Jaccard index, we can infer the extent to which PCM filters out specific types of content or language, potentially leading to more uniform discussions. Figure 3a illustrates the impact of varying PCM levels on community discourse, achieved by adjusting the toxicity score threshold and measuring the Jaccard index at each level. The results indicate a significant shift in word usage within communities, particularly when users' toxicity thresholds are set below approximately 0.1. This observation suggests that lower toxicity thresholds under PCM may substantially narrow the range of expressed views, possibly affecting the diversity and richness of community discussions.

Figure 3b depicts the relationship between the Jaccard index and information loss at various toxicity thresholds. Interestingly, while information loss increases, the filtered content does not significantly impact language use in the community until the information loss exceeds approximately 0.8. This indicates that *AIL* does not notably affect the community's shared reality unless the information asymmetry is substantial.

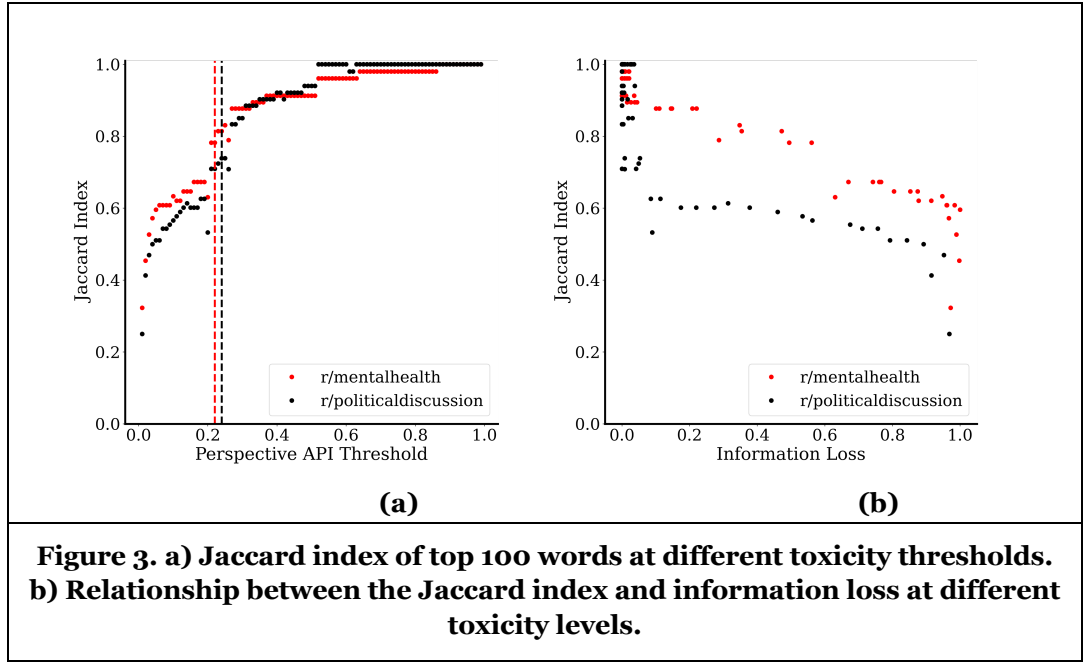

**Figure 3. a) Jaccard index of top 100 words at different toxicity thresholds. b) Relationship between the Jaccard index and information loss at different toxicity levels.**

## Discussion

Our study contributes to understanding the emergent outcomes of Personalized Content Moderation (PCM) on social media platforms. We have shown that PCM leads to asymmetric information loss (AIL), stemming from its personalized approach to content moderation. Our work introduces AIL as a novel concept for analyzing the unintended consequences of PCM. Further analysis in our study indicates that PCM tools may also encourage the formation of echo chambers and filter bubbles, potentially leading to increased community polarization. These findings suggest that PCM can alter the dynamics of online communities on social media platforms.

As users, developers, and scholars engage with Personalized Content Moderation (PCM) tools on social media platforms, they may start considering strategies to mitigate the emergent outcomes of these tools. With the rapid advancement of Large Language Models (LLMs), such as ChatGPT, there is potential for these platforms to offer personalized content modification. This approach could involve modifying content to reduce toxicity levels while retaining the intended meaning of the messages.

## Limitations and Future Research

Our study has certain limitations. Firstly, the measurement of information loss based on the cosine similarity of embeddings could be problematic due to variations in individual interpretations and the inherent subjectivity in understanding messages. Secondly, while our study assumes the application of PCM to existing discussion posts, a more comprehensive observation of PCM's emergent outcomes should consider its effects on the community over time, given the dynamic and evolving nature of online communities. Lastly, accurately identifying echo chambers and filter bubbles necessitates a more advanced analysis of the communities involved.

Our preliminary analysis revealed PCM's emergent outcomes through empirical investigation. However, this analysis necessitates further development and validation of the theoretical hypothesis through these empirical results. As a next step, we aim to develop a comprehensive theoretical framework. By refining the theoretical framework, scholars and developers can better examine these systems and devise interventions to mitigate undesirable emergent outcomes.

Although our preliminary analysis demonstrated emergent outcomes of PCM in online communities. A more extensive empirical investigation is needed. We plan to improve our methodologies using more advanced techniques to analyze AIL, echo chambers, and filter bubbles. In future studies, we also plan to conduct observational studies to examine the effects of PCM on these emergent outcomes.

## Conclusion

Moderation on social media platforms is now more important than ever before. As Personalized Content Moderation (PCM) tools offer user-specific moderation and enhance user experience, it may also impact emergent outcomes on social media platforms. Our study shows that PCM leads to asymmetric information loss (AIL), potentially affecting the development of shared understanding among users. We further demonstrated that PCM could foster the creation of echo chambers and filter bubbles, resulting in increased polarization. Our study is the first to identify the consequences of PCM and to highlight its unintended negative impact on online communities.